\documentclass[11pt]{article} 
\usepackage{moriond,epsfig} 

\newcommand{\imag}{\Im {\rm m}} 
\newcommand{\real}{\Re {\rm e}}

\def\Journal#1#2#3#4{{#1} {\bf #2}, #3 (#4)} 
 
\def\PREP{{\em Phys. Rep.}}

\def\be{\begin{equation}} 
\def\ee{\end{equation}} 
\def\bea{\begin{eqnarray}} 
\def\eea{\end{eqnarray}} 
 
\begin{document} 
\vspace*{4cm} 
\title{CHARGINOS AND NEUTRALINOS AT $e^+e^-$ LINEAR COLLIDERS} 
 
\author{JAN KALINOWSKI } 
 
\address{Institute of Theoretical Physics, Warsaw University\\ 
Ho\.za 69, 00681 Warsaw, Poland} 
 
\maketitle\abstracts{ 
It is shown how the fundamental gaugino and higgsino parameters of the
chargino and neutralino system in the MSSM can completely be
determined in high--precision experiments at $e^+e^-$ linear colliders
even if in their initial phase only the light charginos 
$\tilde{\chi}^\pm_1$ and the light
neutralinos $\tilde{\chi}^0_1$ and $\tilde{\chi}^0_2$ are
kinematically accessible.  }
 
\section{Introduction} 
In the Minimal Supersymmetric Standard Model (MSSM) the light chargino
$\tilde{\chi}^\pm_1$ and the two light neutralinos $\tilde{\chi}^0_1$
and $\tilde{\chi}^0_2$ are expected \cite{Acc} to be among the
lightest supersymmetric particles.  The lightest neutralino
$\tilde{\chi}^0_1$ is commonly assumed to be the lightest SUSY
particle (LSP), stable in the $R$-parity conserving case, and
invisible.  The chargino and neutralino masses are given in terms of
the fundamental parameters in the gaugino/higgsino sector, the U(1)
and SU(2) gaugino masses $M_1$ and $M_2$, the higgsino mass $\mu$, and
the ratio of Higgs vacuum expectation values $\tan\beta$. If
$|\mu|<<|M_1|,\, |M_2|$, the light chargino and light neutralinos are
higgsino-dominated and $m_{\tilde{\chi}^\pm_1} \sim
m_{\tilde{\chi}^0_1} \sim m_{\tilde{\chi}^0_2}$. In the opposite,
gaugino-dominated case, $m_{\tilde{\chi}^\pm_1} \sim
m_{\tilde{\chi}^0_1} \sim 2 m_{\tilde{\chi}^0_2}$ is expected {\it
e.g.} if the gaugino masses unify at high scale in supergravity
inspired scenarios.  Thus ${\tilde{\chi}^\pm_1}$ and $
{\tilde{\chi}^0_2}$ cannot be much heavier than the LSP
${\tilde{\chi}^0_1}$.  Therefore one can envisage a scenario in which
in the initial phase of future $e^+e^-$ colliders, only the
$\tilde{\chi}^\pm_1$, $\tilde{\chi}^0_1$ and $\tilde{\chi}^0_2$ states
will be accessed kinematically with all other supersymmetric particles
being too heavy to be produced. The question then arises, which we
address here, to what extend the fundamental SUSY parameters of the
gaugino/higgsino sector can be reconstructed from the initially
limited experimental input.
 
It has been demonstrated in the literature\cite{CKMZ,1A} that if all
chargino and neutralino states are accessible experimentally, the
measurement of their masses and production cross sections with
polarized $e^+e^-$ beams allows us to derive the parameters $M_1$,
$M_2$, $\mu$ and $\tan\beta$ analytically at tree level.\footnote{For
precision measurements, however, loop corrections will have to be
included bringing in all soft SUSY breaking parameters and ultimately
one will have to rely on a global fit to all experimental data.}  Here
I will report how these parameters can nevertheless be determined with
the limited experimental input available form light charginos and
light neutralinos produced in pairs in $e^+e^-$ collisions
\begin{eqnarray} 
\label{proc} 
e^+e^-\,\rightarrow\,\,\tilde{\chi}^+_1\,\tilde{\chi}^-_1,\hskip 0.5cm  
\tilde{\chi}^0_1\,\tilde{\chi}^0_2  
\end{eqnarray} 
with visible final states.
 
\section{The Chargino System} 
Defining the mixing angles in the unitary matrices diagonalizing the
chargino mass matrix by $\phi_L$ and $\phi_R$ for the left-- and
right--chiral fields, the fundamental SUSY parameters $M_2$, $|\mu|$,
$\cos \Phi_\mu$ and $\tan\beta$ can be derived from the chargino
masses and $c_{2L,2R}=\cos 2\phi_{L,R}$,
\begin{eqnarray} 
M_2&=&m_W \sqrt{\Sigma-\Delta\, (c_{2L}+c_{2R})} \label{eq:m2} \\ 
|\mu|&=& m_W \sqrt{\Sigma+\Delta\, (c_{2L}+c_{2R})}\label{eq:mu}\\ 
\cos\Phi_\mu&=&\frac{\Delta^2(2-c^2_{2L}-c^2_{2R}) 
                  -\Sigma}{ 
         \sqrt{[1- \Delta^2 (c_{2L}-c_{2R})^2]\, 
               [\Sigma^2-\Delta^2(c_{2L}+c_{2R})^2]}} 
	       \label{eq:cosphi}\\ 
\tan\beta&=&\sqrt{[1-\Delta (c_{2L}-c_{2R})]/ 
                       [1+\Delta (c_{2L}-c_{2R})]}\label{eq:tanb} 
\end{eqnarray} 
where $\Sigma = (m^2_{\tilde{\chi}^\pm_2}+m^2_{\tilde{\chi}^\pm_1} - 2
m^2_{_W}) /2 m^2_{_W}$ and $\Delta =
(m^2_{\tilde{\chi}^\pm_2}-m^2_{\tilde{\chi}^\pm_1})/ 4 m^2_{_W}$.  The
$\cos 2\phi_{L,R}$ can be determined uniquely\cite{SONG,6A} from the
measurement of cross sections\footnote{The cross sections for
processes (\ref{proc}) depend on the sneutrino and selectron masses
which we assume, for the sake of simplicity, to be measured
elsewhere.}
$\sigma(e^+e^-\,\rightarrow\,\,\tilde{\chi}^+_1\,\tilde{\chi}^-_1)$ at
one energy with polarized beams including transverse beam
polarization, or else if only longitudinal beam polarization is
available, they are measured at two different incoming energies.

If only the light charginos $\tilde{\chi}^\pm_1$ can be produced,  
the mass of ${\tilde{\chi}^\pm_2}$ remains unknown. Then  it depends on the  
CP properties of the higgsino sector whether the parameters of 
eqs.(\ref{eq:m2}--\ref{eq:tanb})  can be determined or  
not in the chargino system alone. \\ 
{\it (i)} If $\mu$ is real, as suggested by the electric dipole 
moments, eq.(\ref{eq:cosphi}) can be exploited to 
determine $m^2_{\tilde{\chi}^\pm_2}$ from $\cos\Phi_\mu =\pm 1$ up to 
at most a two--fold ambiguity.\cite{SONG,6A}   \\  
{\it (ii)} If $\mu$ is complex, the parameters in 
eqs.(\ref{eq:m2}--\ref{eq:tanb}) cannot be determined  
in the chargino sector alone. 
However, they can be calculated as functions of the unknown heavy 
chargino mass $m_{\tilde{\chi}^\pm_2}$. Actually, there are two 
classes of solutions corresponding to the two values $\Phi_\mu$ and 
$(2\pi -\Phi_\mu)$ for the phase of the higgsino mass parameter, {\it 
i.e.}  the sign of $\sin\Phi_\mu$.\footnote{This phase ambiguity can be 
resolved 
by measuring  the sign of CP--odd observables, {\it e.g.}  associated 
with normal 
$\tilde{\chi}^0_2$ polarization in $\tilde{\chi}^0_1 \tilde{\chi}^0_2$ 
pair production.\cite{WYSONG}}  
Although the heavy chargino mass is still unknown,  
its range is bounded from above by eq.(\ref{eq:tanb}),   
and from below 
by not observing the heavy chargino in mixed light$-$heavy pair production 
\begin{eqnarray} 
     {\textstyle{\frac{1}{2}}} \sqrt{s} - m_{\tilde{\chi}^\pm_1}  
\,\leq\, m_{\tilde{\chi}^\pm_2}  
\,\leq\, ({\, m^2_{\tilde{\chi}^\pm_1} 
           +\, 4 m^2_W/| \cos 2\phi_L-\cos 2\phi_R|})^{1/2} 
\end{eqnarray} 
To resolve the two--fold 
ambiguity in case {\it (i)}, and to fix the heavy chargino mass    
in case {\it (ii)} other observables are needed.   
It is interesting to note that in solving both cases  the 
mixed--pair $\tilde{\chi}^0_1 \tilde{\chi}^0_2$ production process can 
be used since at the same time the U(1) gaugino mass parameter $M_1$ 
can be determined.

\section{The Neutralino System} 
The symmetric neutralino mass matrix  is diagonalized by 
a unitary matrix, defined such that the mass eigenvalues  
of the four Majorana fields $\tilde{\chi}^0_i$ 
are positive. The squared mass eigenvalues of the neutralinos are   
solutions of the characteristic equations \cite{CKMZ} 
\begin{equation} 
m_{\tilde{\chi}^0_i}^8-a\, m_{\tilde{\chi}^0_i}^6 
+b\, m_{\tilde{\chi}^0_i}^4-c\, m_{\tilde{\chi}^0_i}^2+d=0 \quad   
{\rm for} \quad i=1,2,3,4 
\label{eq:characteristic} 
\end{equation} 
with the invariants $a$, $b$, $c$ and $d$ given by the  
gaugino mass parameters $M_2$ and $M_1$, and the higgsino 
mass parameter $\mu$, {\it i.e.} the moduli $M_2$, $|M_1|$, $|\mu|$ and the 
phases $\Phi_1$, $\Phi_\mu$. Since   
$a$, $b$, $c$ and $d$  are binomials of $\real{M_1}=|M_1|\,\cos\Phi_1$  
and $\imag{M_1}=|M_1|\,\sin\Phi_1$, the characteristic  
equation (\ref{eq:characteristic}) for each neutralino mass  has the form 
\begin{eqnarray} 
(\real{M_1})^2+(\imag{M_1})^2+ u_i\, \real{M_1}+ v_i\, \imag{M_1} 
 = w_i \quad {\rm for}\quad i=1,2,3,4 
\label{eq:Mphase} 
\end{eqnarray} 
The coefficients $u_i$, $v_i$ and $w_i$ are functions of the  
parameters $M_2$, $|\mu|$, $\Phi_\mu$, $\tan\beta$ and the mass 
eigenvalue $m^2_{\tilde{\chi}^0_i}$ for fixed $i$. The coefficient $v_i$  
is necessarily proportional to $\sin\Phi_\mu$ because physical neutralino 
masses are CP--even: $\sin\Phi_\mu=0$ implies a sign ambiguity in the  
CP--odd quantity 
$\imag M_1$, {\it i.e.} in $\sin\Phi_1$. 
Therefore the  characteristic equation (\ref{eq:Mphase}) defines a circle 
in the $\{\real M_1, \imag M_1\}$ plane for each neutralino mass  
$m_{\tilde{\chi}^0_i}$. With two known light neutralino masses  
$m_{\tilde{\chi}^0_1}$ and $m_{\tilde{\chi}^0_2}$, we  
have two circles which cross at two points, see left panel of Fig~1.  
 
Now we are in position to solve cases {\it (i)} and {\it (ii)} 
discussed above.\\ 
{\it (i)} For $\mu$ real, the two--fold ambiguity for $M_2$, $\mu$ and 
$\tan\beta$ can be resolved by checking which combination provides two 
crossing circles, {\it i.e.} is consistent with the measured neutralino 
masses and production cross section.  
At the same time $|M_1|$ is determined with a remaining two--fold 
ambiguity for  $\imag M_1$.\\ 
{\it (ii)} For $\mu$ complex, the position of two circles in the 
$\{\real M_1, \imag M_1\}$ plane, and therefore their crossings, will 
migrate as functions of the unknown heavy chargino mass, see right panel 
of Fig.~1.  By comparing the predicted with the measured pair--production 
cross sections $\sigma_L\{\tilde{\chi}^0_1\tilde{\chi}^0_2\}$ and 
$\sigma_R\{\tilde{\chi}^0_1\tilde{\chi}^0_2\}$, a unique solution, for 
both the parameters $m_{\tilde{\chi}^\pm_2}$ and $\real M_1, \imag 
M_1$ can be found.  As a result, the additional measurement of the 
cross sections leads to a unique solution for $m_{\tilde{\chi}^\pm_2}$ 
and subsequently to a unique solution for $M_1$, $M_2$, $\mu$ and  
$\tan\beta$ (assuming that the discrete CP ambiguity is 
resolved, see footnote $c$). 
\begin{figure}[tbh] 
\begin{center} 
\vspace*{5mm} 
 \epsfig{file=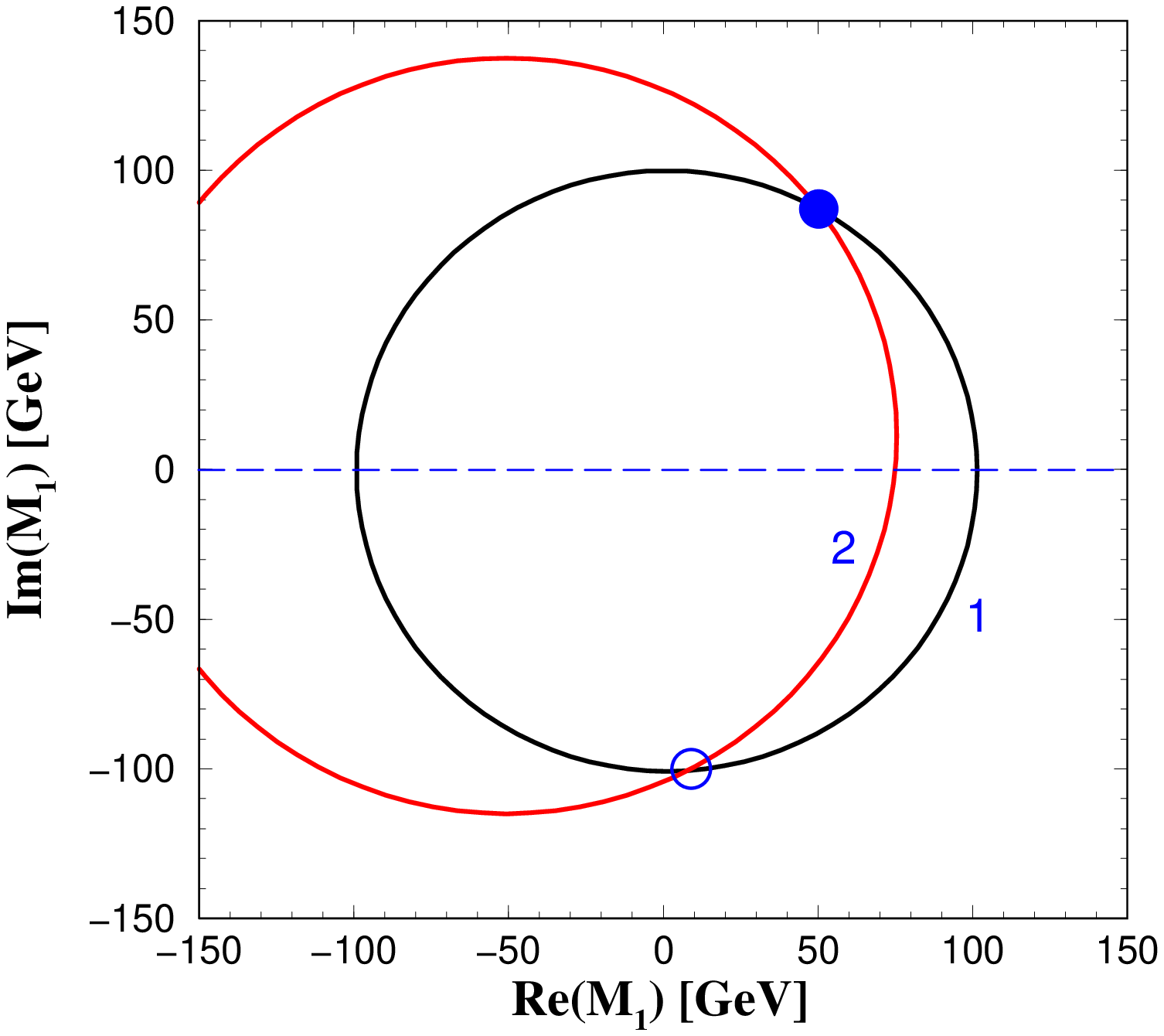,height=6cm,width=6cm} \hspace{6mm}  
 \epsfig{file=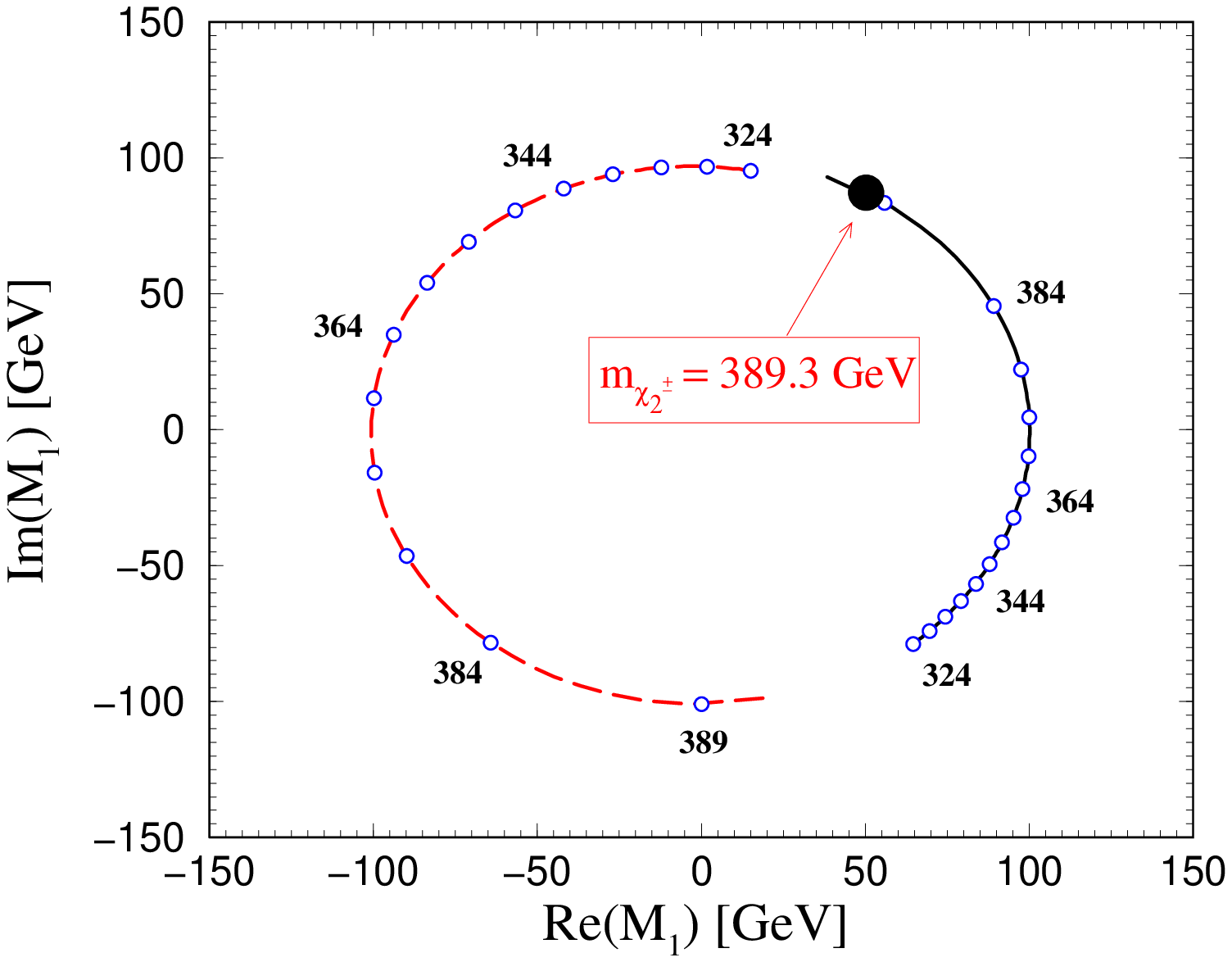,height=6cm,width=6cm} 
\caption{Left: {\it The contours of two measured neutralino masses 
           $m_{\tilde{\chi}^0_1}$ and $m_{\tilde{\chi}^0_2}$ in the 
	   $\{{\real M_1},\,{\imag M_1}\}$ plane. The parameter set 
	   $\{M_2=190.8\, {\rm GeV};\, |\mu|= 365.1\, {\rm GeV},  
	   \Phi_\mu =\pi/8;$ $\tan\beta=10\}$, corresponding to  the point 
(\ref{eq:parameter}), is assumed to be 
	   known from the chargino sector.} 
 Right: {\it Migration of the 
crossing points in the 
	   $\{{\real M_1},\,{\imag M_1}\}$ plane, as functions of the heavy 
	   chargino mass (the small open  
	   circles are spaced by 5 GeV).  
	   The unique solution is determined by the measurement of the  
	   pair--production cross sections (marked by the black dot).} 
           From ref.\protect\cite{CKMZ}.  } 
\label{fig:circle} 
\end{center} 
\end{figure} 
%

\section{Extracting the Fundamental Parameters} 
The strategy described above is illustrated in Fig.~1.  It has been
worked out\cite{CKMZ} for a single reference point for a CP
non--invariant extension of the MSSM, compatible with all experimental
constraints,
%
\begin{eqnarray} 
{\sf RP}:\, 
  (|M_1|, M_2, |\mu|;\, \Phi_1, \Phi_\mu;\, \tan\beta) 
             =(100.5\,{\rm GeV},\, 190.8\, {\rm GeV},\, 
                 365.1\, {\rm GeV};\,\frac{\pi}{3},\,\frac{\pi}{8}; 
		\, 10) 
\label{eq:parameter} 
\end{eqnarray} 
%
These fundamental parameters generate the following light chargino and  
neutralino masses,\\ 
\centerline{ 
$m_{\tilde{\chi}^\pm_1}=176.0$ GeV,   
~~~$m_{\tilde{\chi}^0_1}=98.7$ GeV,   
~~~$m_{\tilde{\chi}^0_2}=176.3$ GeV}  
while the heavy masses are given by\\ 
\centerline{ 
$m_{\tilde{\chi}^\pm_2}=389.3$ GeV,   
~~~$m_{\tilde{\chi}^0_3}=371.8$ GeV,   
~~~$m_{\tilde{\chi}^0_4}=388.2$ GeV}  
The cross sections depend on the sneutrino and selectron masses which we 
assume, for the sake of simplicity, to be measured in threshold scans\\ 
\centerline{    
$m_{\tilde{\nu}_{_L}}=192.8$ GeV,    
~~~$m_{\tilde{e}_{_L}}=208.7$ GeV,  
~~~$m_{\tilde{e}_{_R}}=144.1$ GeV}  
The cross sections for chargino and neutralino pair--production with polarized 
beams, 
\begin{eqnarray} 
&& \sigma_L\{\tilde{\chi}^{\!+}_1\tilde{\chi}^{\!-}_1\!\} = 679.5\, {\rm fb}  
   \qquad \ \ 
   \sigma_R\{\tilde{\chi}^{\!+}_1\tilde{\chi}^{\!-}_1\!\} = 1.04\, {\rm fb}  
   \label{eq:c11 x-section}\\[1mm] 
&& \sigma_L\{\tilde{\chi}^0_1\tilde{\chi}^0_2\} = 327.9\, {\rm fb}  
   \qquad \ \ 
   \sigma_R\{\tilde{\chi}^0_1\tilde{\chi}^0_2\} = 16.4\, {\rm fb}  
   \label{eq:n12 x-section} 
\end{eqnarray} 
at $\sqrt{s}=500$ GeV are sufficiently large (between $\sim 7\times
10^5$ and $1\times 10^3$ events for $\tilde{\chi}^+_1\tilde{\chi}^-_1$
and $\tilde{\chi}^0_1\tilde{\chi}^0_2$ can be expected at
TESLA\cite{TDR}), allowing the analysis of the properties of the
chargino $\tilde{\chi}^\pm_1$ and the neutralinos
$\tilde{\chi}^0_{1,2}$ with high precision.

{\it To summarize.} If only the light chargino  
and the two light neutralinos  
can be accessed kinematically in the initial phase of $e^+e^-$ linear  
colliders, measurements of their masses and production cross sections 
with polarized beams,   
and $\tilde{\chi}^0_2$ polarization in the process  
$e^+e^-\rightarrow \tilde{\chi}^0_1 \tilde{\chi}^0_2$,  
allow us to perform a complete and precise analysis of 
the basic MSSM parameters in the gaugino/higgsino sector: $(M_1, M_2; \mu; 
\tan\beta)$. For more details and references, see 
ref.~\cite{CKMZ}.

\section*{Acknowledgments} 
I would like to thank Organisers for inviting me to Moriond and 
for financial support.    I am grateful to S.Y. Choi, G. Moortgat--Pick 
and P.M. Zerwas for many stimulating discussions. 
Work supported in part by the KBN Grant No.  
5 P03B 119 20 (2001-2002).

\end{document}